\documentclass[apj]{emulateapj}
\usepackage{amssymb}
\usepackage{multirow}
\usepackage{graphicx}
\usepackage{subfigure}

\shorttitle{}
\shortauthors{Webster, Sutherland \& Bland-Hawthorn}

\newcommand{\comments}[1]{}

\begin{document}

\title{Star Formation in Ultrafaint Dwarfs: Continuous or Single-age Bursts?}

\author{David Webster}
\affil{Sydney Institute for Astronomy, School of Physics, University of Sydney, NSW 2006, Australia}

\author{Joss Bland-Hawthorn}
\affil{Sydney Institute for Astronomy, School of Physics, University of Sydney, NSW 2006, Australia}

\author{Ralph Sutherland}
\affil{Research School of Astronomy \& Astrophysics, Australian National University, Cotter Rd,
Weston, ACT 2611, Australia}

\email{d.webster@physics.usyd.edu.au}

\begin{abstract}
We model the chemical evolution of six UFDs: 
Bootes I, Canes Venatici II, Coma Berenices, Hercules, 
Leo IV and Ursa Major I, based on their recently determined star formation histories. We show that two single-age bursts 
cannot explain the observed [$\alpha$/Fe] vs [Fe/H] distribution in these galaxies and that some 
self-enrichment is required within the first burst. An alternative scenario is modelled, in which star 
formation is continuous except for short interruptions when one or more supernovae temporarily blow the dense gas out from the centre of the system. 
This model allows for self-enrichment and can reproduce the chemical abundances of the UFDs in which the second burst is only a trace population. 
We conclude that the most likely star formation history is one or two extended periods of star formation, with 
the first burst lasting for at least 100~Myr. As found in earlier work, the observed properties of UFDs can be explained by formation at a low 
mass ($M_{\rm{vir}}\sim10^7$~M$_\odot$), rather than being stripped remnants of much larger systems.

\end{abstract}


\section{Introduction}

The discovery of very faint low-mass galaxies, known as ultra-faint dwarfs ($L<10^5$~L$_\odot$) has provided a low-redshift 
method of investigating the formation and evolution of the early baryonic systems. UFDs contain only old, metal-poor 
stellar populations and may retain relatively unpolluted chemical signatures of the first generations of stars. Recent work has made progress in 
determining the star formation history \citep{weisz14,brown14} and chemical abundances \citep{frebel12,gilmore13,vargas13} of these systems. 

For most stars in UFDs, only medium-resolution spectroscopy is available, however this is sufficient to determine iron and $\alpha$ element 
abundances \citep{kirby09,kirby10}. This allows use of the idea of \citet{tinsley79}, who suggested that the 
enhanced [$\alpha$/Fe] ratio in halo stars could be explained by the time delay between Type II and Type Ia supernovae. The 
progenitors of Type Ia supernovae are evolved low and 
intermediate-mass stars with longer lifetimes than the massive stars that are the progenitors of Type II supernovae, such that there is a $\sim100$~Myr 
period after star formation commences in a galaxy when only Type II supernovae enrich the gas. Type II supernovae eject much more $\alpha$ elements 
relative to iron than Type Ia supernovae, resulting in enhanced [$\alpha$/Fe] for stars that form in the first 100~Myr of star formation in a system. 
A galaxy that only contains stars with [$\alpha$/Fe] above or close to the mean of 0.35 expected for Type II supernovae in a typical initial mass 
function \citep[e.g.][]{salpeter55,kroupa01} is likely to have formed stars for less than 100~Myr \citep{cayrel04,frebel12}. 

Enrichment from Type Ia supernovae is not the 
only possible explanation for low [$\alpha$/Fe], which can also be explained by an initial mass function (IMF) that favours lower-mass Type II 
supernovae, or by types of 
supernovae in which most $\alpha$ elements do not escape into the interstellar medium \citep{karlsson12}. Observationally 
disentangling the contribution of the decline in [$\alpha$/Fe] as a result of Type II supernovae as compared to Type Ia supernovae is 
therefore difficult and is likely to require high-resolution observations of iron-peak and s-process elements. It should also be noted that 
there is not necessarily a direct relationship between [Fe/H] and time, as the merging of gas clouds can create multiple metallicity populations 
\citep{wise12}.

The low luminosity of the UFDs means that chemical abundance data is available for only a few stars in each galaxy, 
resulting in large uncertainties for individual systems. yHowever, the situation is improving, with \citet{brown14} 
providing [Fe/H] for a large sample of stars in six UFDs and \citet{vargas13} 
determining [$\alpha$/Fe] abundances for 61 stars in eight UFDs. The [Fe/H] distribution from these observations was then used to determine the ages of 
the stars relative to the M92 globular cluster.

\citet{brown14} used isochrone fitting to determine the star formation history (SFH) of six UFDs; Bootes 1, Coma Berenices, Canes Venatici II, Hercules, 
Leo IV and Ursa Major I. Without the constraint of spectroscopic abundances for [$\alpha$/Fe], they found that the SFH could be fit by a two single-age burst model with three parameters: 
the ages of the two components and the proportion of stars in each burst. Adding parameters for the duration of the two bursts did not improve the fit, 
which the authors suggest indicates a narrow age range for the stars within each burst. 

In this work we simulate chemical abundances given two possible 
star formation histories using the models of $M_{\rm{vir}} = 10^7$~M$_\odot$ halos presented in \citet{webster14} \& \citet{hawthorn15}. \citet{collins14} 
showed that Bootes I and Hercules have circular velocities of 5-7~kms$^{-1}$, consistent with halo masses this low. We also investigate whether 
observed chemical abundances \citep{gilmore13,vargas13,brown14} are consistent with the \citet{brown14} star formation histories. 

\section{Models}

The simulations used to model the chemical abundances are described in \citet{webster14} \& \citet{hawthorn15}. The 3D hydro/ionization code \emph{Fyris Alpha} 
\citep{sutherland2010} was used to model the effects of a 25~M$_\odot$ star on gas in an $M_{\rm{vir}} = 10^7$~M$_\odot$ halo. The 
density and metallicity distribution of the gas after the first supernova was then used as a template to simulate the effects of later supernovae, allowing 
an estimation of star formation and gas enrichment over periods of up to 600~Myr. Using the method of \citet{argast00}, a number of cells
 were randomly selected, with stars forming in these cells with a probability proportional to the square of the density of the gas. 

In our model of two single-age bursts, the first burst stars form in gas enriched from [Fe/H]~$=-4$ by only a single 25~M$_\odot$ star. 
The density and 
metallicity distribution for the first burst gas is taken from the hydrodynamical model 15~Myr after the supernova. At this time most of the enriched gas 
has returned to the centre of the galaxy, such that the number of cells with $n_H>10$~cm$^{-3}$ is $>70\%$ of the number in the undisturbed state. 
The second burst forms in gas enriched by Type II and Type Ia supernovae from the first burst of stars, with yields as in \citet{woosley95} \& \citet{iwamoto99}. 
The number of Type Ia supernovae is similar to the number of Type II supernovae. Bootes I is enriched 
by only Type II supernovae, because the two bursts in \citet{brown14} are only 100~Myr apart. The continuous model is described in 
\citet{webster14}. The gas is enriched by only Type II supernovae for the first 100~Myr, after which Type Ia 
supernovae occur with a rate as in \citet{jiminez14}. Because higher mass stars yield more alpha elements, varying the mass of the initial star would alter [$\alpha$/Fe] at low [Fe/H], 
meaning that we should not necessarily expect the models to fit the observations at low [Fe/H].  
However, this effect is washed out after a few supernovae.

\begin{figure*}
	\centering
	\includegraphics[width=.9\textwidth]{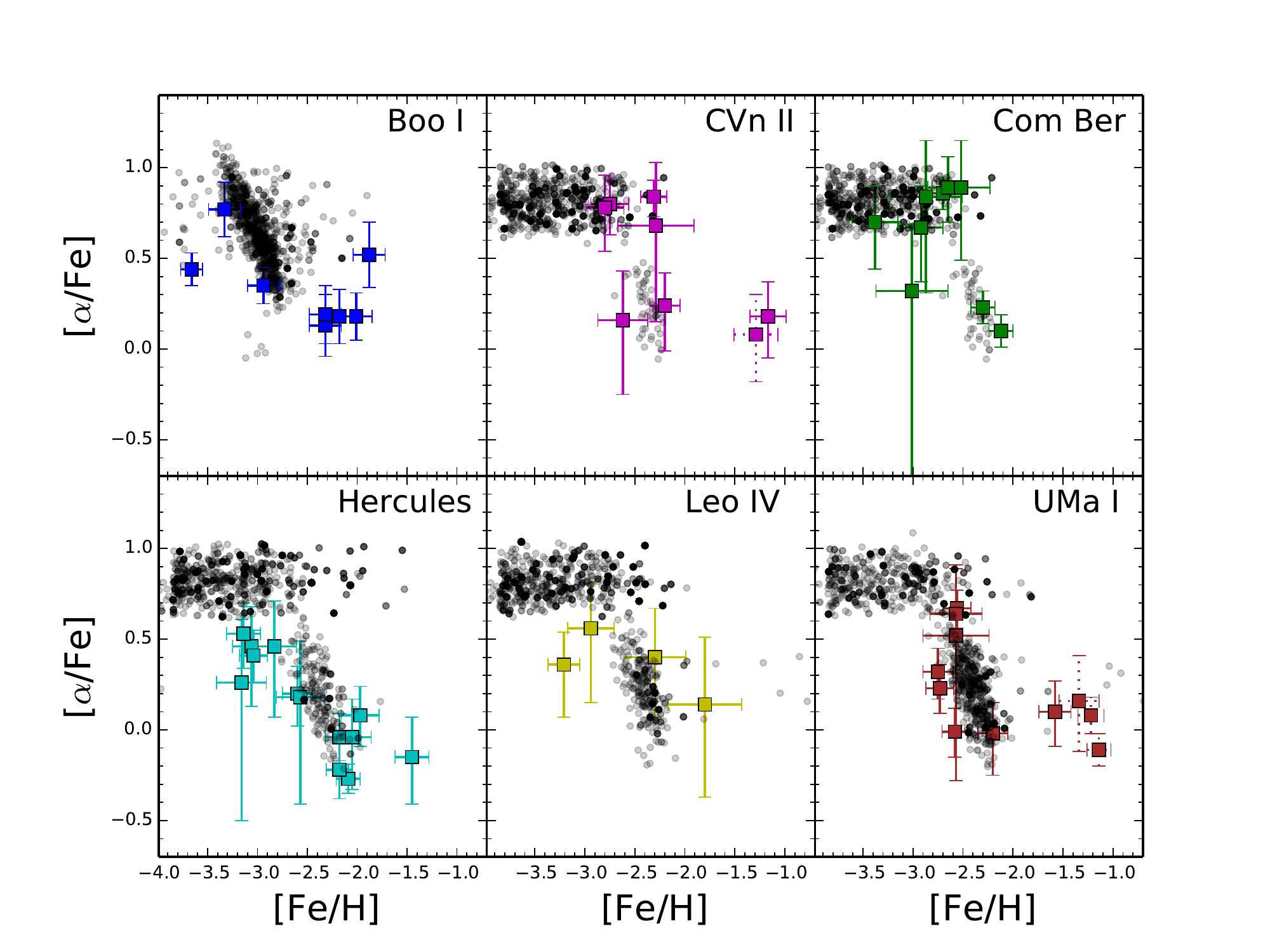}
	\caption{[$\alpha$/Fe] vs [Fe/H] for 6 UFDs from our model (black points) and observations (coloured squares) from \citet{gilmore13} (Bootes I) \& \citet{vargas13} (the other five galaxies). Dotted error bars represent stars whose membership of the system is in doubt (J. Simon, private comm.)}
	\label{f:afefeh}
\end{figure*}

\begin{figure*}
	\centering
	\includegraphics[width=.9\textwidth]{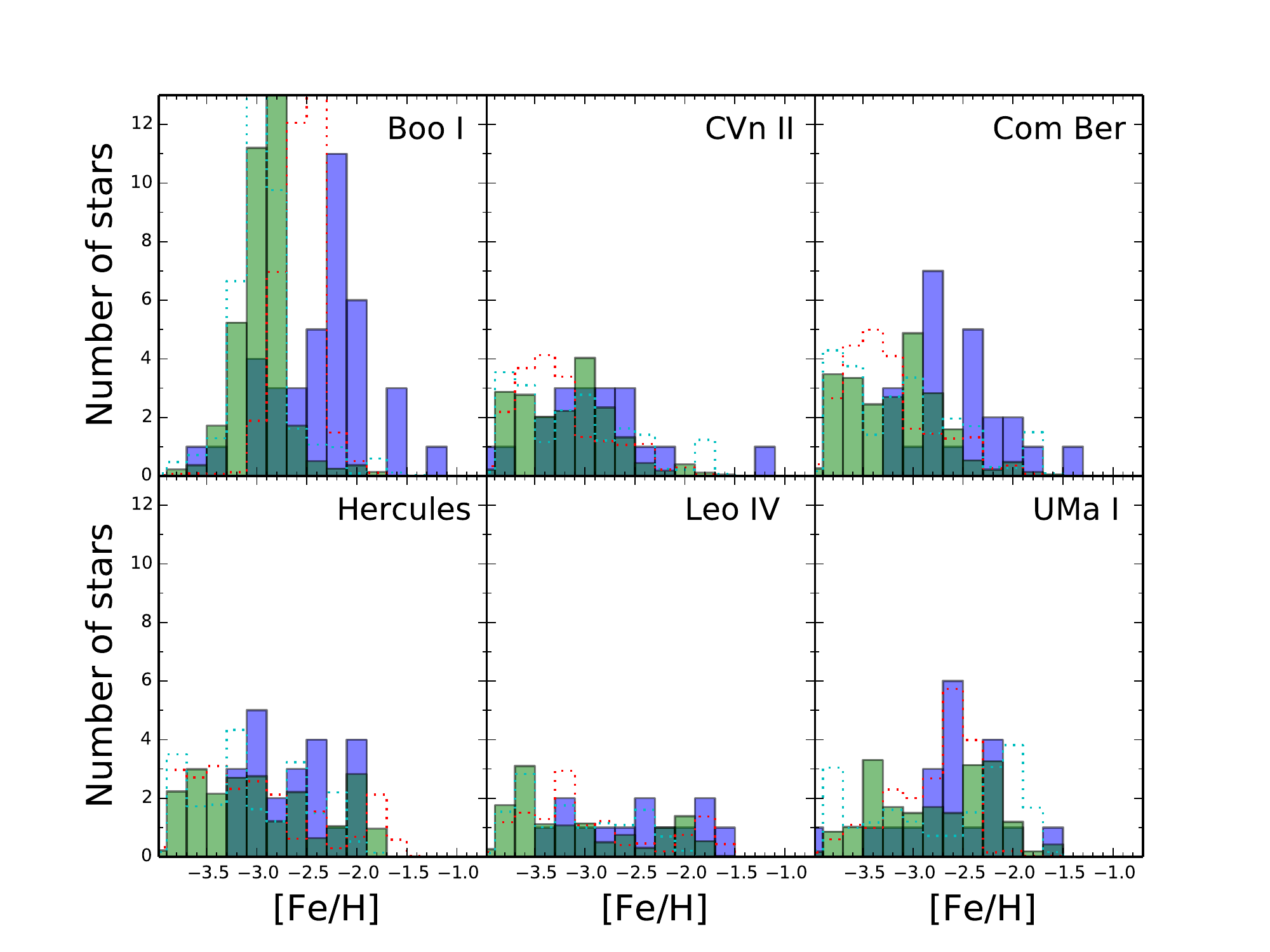}
	\caption{Histograms of [Fe/H] from our model (green) and the \citet{brown14} observations (blue) for 6 UFDs. The modelled histogram is normalised to the number of observed stars in each UFD. The dotted histograms correspond to a model with a higher Type II SN rate and lower Type Ia SN rate (red), and a lower Type II SN rate and higher Type Ia SN rate (light blue).}
\label{f:fehhist}
\end{figure*}

\subsection{Single-age bursts}

Based on colour-magnitude diagrams, \citet{brown14} modelled the star formation histories of six UFDs as two single-age populations, determining the ages of the bursts and 
the proportion of stars in each burst. A model with extra parameters for the duration of star formation in each burst did not improve their fit, 
suggesting a narrow age range within each burst.


\subsubsection{Bootes I}

The two-burst model of \citet{brown14} has 97\% of the stars in Bootes I forming in the second burst, with the 
two bursts being only 100~Myr apart. This is the only galaxy for which the \citet{brown14} model favours such closely spaced bursts, as well as the only 
case where significantly more stars form in the second burst. 26 of the 38 
observed stars (68\%) in Bootes I have [Fe/H]~$>-2.5$, compared to $\lesssim50\%$ for the other five galaxies. Our model for Bootes Iis shown in the first panel of Figs.~\ref{f:afefeh} and \ref{f:fehhist}, along with the [Fe/H] data from \citet{brown14} and [$\alpha$/Fe] 
data from \citet{gilmore13}. In our model, the 3\% of stars that form in the first burst enrich the gas to [Fe/H]~$\sim-3$, with a spread in [$\alpha$/Fe] 
resulting from variations 
in the extent to which these stars enrich different regions of the galaxy. 

A cross-correlation between the observed [Fe/H] histogram and our modelled histogram 
gives a lag of 0.6~dex for the model. This could be explained by a higher initial metallicity in Bootes I than in our model, or a top-heavy IMF 
such that a greater proportion of the stars in the first burst 
produced supernovae. The red dotted histogram, corresponding to an increased Type II SN rate, shows a much smaller lag compared to the observations.

[$\alpha$/Fe] observations from 
\citet{gilmore13} indicate at least some self-enrichment, with evidence of a decline in [$\alpha$/Fe] with increasing [Fe/H]. This could be 
explained by Type Ia supernovae, but can also be explained by lower mass Type II supernovae, with stochastic sampling of a Kroupa IMF selecting 
many 8-15~M$_\odot$ star in Bootes I, reducing [$\alpha$/Fe]. If Type II supernovae are the cause of the decline, the \citet{brown14} scenario 
with two single-age bursts 100~Myr apart is possible for Bootes I.

\subsubsection{Canes Venatici II}

The \citet{brown14} model found that 95\% of the stars in Canes Venatici II formed in the first burst. 
Our modelled [$\alpha$/Fe] and [Fe/H] for this scenario is shown along with observational 
data \citep{vargas13,brown14} in the second panel of Figs.~\ref{f:afefeh} \& \ref{f:fehhist}. 
While our model of the two-burst scenario can fit most of the stars, it produces insufficient scatter to fit the stars at [Fe/H]~$\sim-1.2$. Furthermore, four of the eight stars for which [$\alpha$/Fe] data is available show [$\alpha$/Fe] suppressed by 0.5~dex compared to the other four stars. 
If low [$\alpha$/Fe] reflects self-enrichment, these stars must have formed in the second burst. However, \citet{brown14} suggest that 95\% of the 
stars in Canes Venatici II formed in the first burst. Selecting eight stars randomly from such a distribution gives a probability of $3\times10^{-4}$ 
that four or more will be second burst stars. Removing one of the high-metallicity stars for which the membership of the system is uncertain (J. Simon, private comm.) 
increases this probability to $4\times10^{-3}$. As with Bootes I, the cross-correlation showed a lag of 0.6~dex in [Fe/H] for the model compared to the observations.


\subsubsection{Coma Berenices}

Coma Berenices shows a lag of 0.4~dex in [Fe/H] for the best-fitting model compared to the observations. 
As shown in Fig.~\ref{f:afefeh}, two of the nine stars for which [$\alpha$/Fe] is available show suppressed [$\alpha$/Fe] abundances, suggesting that they belong to the second burst. Using the same method as for Canes Venatici II, 
this is reasonably unlikely ($p=0.05$) given that \citet{brown14} determine that 96\% of the stars belong to the first burst. If the two-burst model is to explain Coma Berenices and Canes Venatici II, it requires more stars in the 
second burst, or enrichment within the first burst. 

\subsubsection{Hercules}

The \citet{brown14} model for Hercules has 82\% of the stars in the first burst. As with the previous systems, the model overestimates the number of low-metallicity stars, with the cross-correlation showing a lag of 0.6~dex 
in [Fe/H] for the model. Fig.~\ref{f:afefeh} shows that 
[$\alpha$/Fe] observed in Hercules is lower than predicted by the model, suggesting that the gas that formed the first burst of stars was 
enriched by stars with lower alpha abundances than a 25~M$_\odot$ supernova. Five of the 13 stars show sub-solar [$\alpha$/Fe], suggesting 
Type Ia enrichment. This is possible, but unlikely ($p=0.07$) given that the \citet{brown14} model has 82\% of the stars in the first burst. 

\subsubsection{Leo IV}

Limited abundance data is available for Leo IV, which contains only four stars with known [$\alpha$/Fe] 
\citep{vargas13} and 13 with [Fe/H] \citep{brown14}. The four stars with [$\alpha$/Fe] abundances are 
consistent with no decline with increasing [Fe/H] or a slight decline. There is 
insufficient data to conclude whether Leo IV is consistent with the \citet{brown14} two-burst model. 

\begin{figure*}
	\centering
	\includegraphics[width=.9\textwidth]{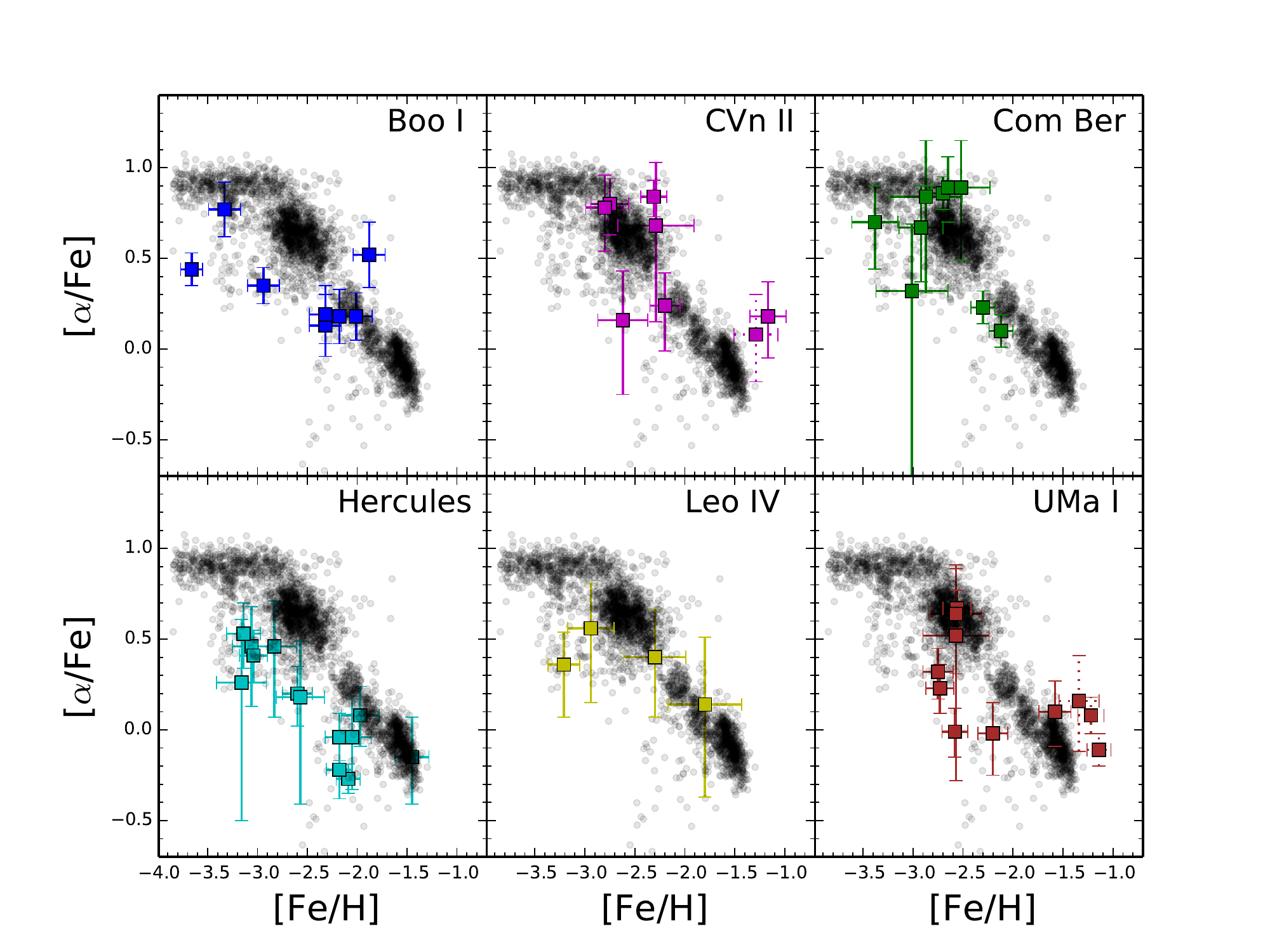}
	\caption{[$\alpha$/Fe] vs [Fe/H] in our continuous model (black points) and as observed \citep{gilmore13,vargas13} (coloured squares). Dotted error bars represent stars whose membership of the system is in doubt (J. Simon, private comm.).}
	\label{f:afeconthist}
\end{figure*}

\begin{figure*}
	\centering
	\includegraphics[width=.9\textwidth]{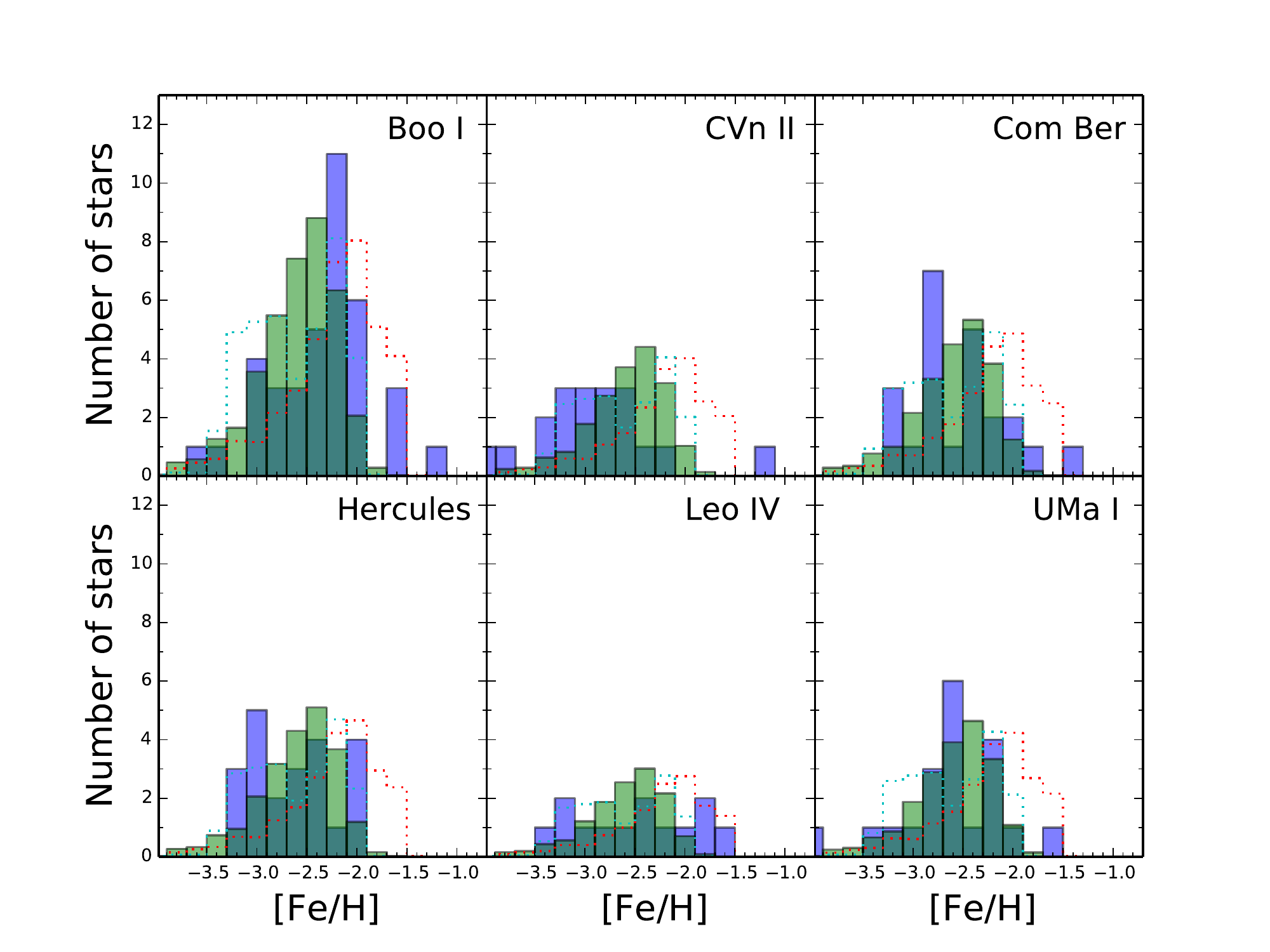}
	\caption{[Fe/H] histograms for stars in our continuous model (blue) and the \citet{brown14} observations (green). The dotted histograms correspond to a model with a higher Type II SN rate and lower Type Ia SN rate (red), and a lower Type II SN rate and higher Type Ia SN rate (light blue).}
	\label{f:fehconthist}
\end{figure*}

\subsubsection{Ursa Major I}

Ursa Major I has half-light mass of 2.6$^{+1.2}_{-1.1}\times10^7$~M$_\odot$ \citep{collins14} and a 
luminosity of $1.4\pm0.4\times10^4$~L$_\odot$, which is much more massive and luminous than our model and therefore the results from the model 
 should be treated with caution. It is the only galaxy studied by \citet{brown14} with similar numbers of stars forming in each burst. 
As shown in Fig.~\ref{f:afefeh}, the model can explain the observed abundances except for the stars with [Fe/H]~$\sim-1.5$. However, three of these 
higher-metallicity stars are included in \citet{vargas13} but not in \citet{brown14} because they are no longer believed to be members (J. Simon, private comm.). When these three stars are excluded, the model is a much better fit to Ursa Major I.

\subsubsection{Summary}

The two single-age burst model of \citet{brown14} found that four of the six UFDs studied formed 75-96\% of stars in the first burst. 
However, 38\% (13/34) of stars in these systems show [$\alpha$/Fe] suppressed $\gtrsim$0.5~dex compared to the high [$\alpha$/Fe], low [Fe/H] 
stars in the same galaxy. These stars show [$\alpha$/Fe] $\lesssim0.2$, indicative of enrichment either from Type Ia supernovae or lower mass 
Type II supernovae 
after the time at which the high [$\alpha$/Fe] stars formed. To fit observations of [$\alpha$/Fe], the two single-age bursts model requires 
a more even distribution of stars between the first and the second burst.


\subsection{Continuous model}

In \citet{webster14}, we presented a star formation history where each supernova in a $10^7$~M$_\odot$ galaxy temporarily blows out 
the gas from the centre of the system, pausing star formation for $\sim10-15$~Myr until the gas recovers. Star formation proceeds 
in the intervals between the supernovae, which are typically $\sim$10-20~Myr, but can be as long as 30-40~Myr. Longer, possibly permanent 
pauses may be caused by a large number of supernovae occurring nearly simultaneously. The output from this model is presented in Figs.~\ref{f:afeconthist} 
\& \ref{f:fehconthist} along with observed abundances from \citet{brown14,vargas13,gilmore13}.

Like the two-burst model, the continuous model shows a clump of stars at low [Fe/H] with [$\alpha$/Fe]~$\sim$0.8, however this is a smaller proportion of stars than for the bursts model. The gas is then enriched by Type II supernovae, resulting in stars forming with 
[Fe/H]~$\sim-2.5$ and [$\alpha$/Fe]~$\sim0.6$. This level of [$\alpha$/Fe] is higher than the average of $\sim0.35$ from Type II supernovae because 
the low star formation rate results in a low supernova rate, meaning that the enhanced $\alpha$ abundances present in the initial gas and from the 
25~M$_\odot$ star have not yet been washed out. After 100~Myr, Type Ia supernovae enrich the gas, eventually reducing [$\alpha$/Fe] to sub-solar levels. 

The [Fe/H] histogram for this model produces significantly more higher metallicity stars than the two single-age bursts due to the self-enrichment 
resulting from extended star formation. The number of stars at each metallicity 
increases until [Fe/H]~$=~-2.5$, as expected for a reasonably constant rate of enrichment, because [Fe/H] is a logarithmic scale. 
There is then a decline caused by the start of Type Ia enrichment. Type Ia supernovae yield much more Fe than Type II, resulting in a more rapid 
increase in [Fe/H] and therefore fewer stars at a given metallicity. The number of stars at each [Fe/H] then begins to increase again, followed by a decline due to the truncation of star formation. While the observed [Fe/H] histograms contain limited stars, there may be signs of this pattern in 
Coma Berenices, Ursa Major I \& Hercules, all of which show valleys near [Fe/H]~$=~-2.5$, which is the metallicity at which these 
systems start to show evidence of Type Ia supernovae. 

\subsection{Comparison between the models}

To compare the models we implement a method based on Section 3.2 of \citet{price14}. For each point in the model, the probability of a modelled star $k$ matching the observed star $s$ is:

\begin{equation}
p_{ks} = \frac{1}{2\pi\sigma_x\sigma_y}\rm{exp}(-(\frac{(x_{\rm{mod}}-x_{\rm{obs}})^2}{2\sigma^2_x}+\frac{(y_{\rm{mod}}-y_{\rm{obs}})^2}{2\sigma^2_y}))
\end{equation}

where x = [Fe/H] and y = [$\alpha$/Fe]. This is then summed over all the modelled stars and normalised by the number of stars in the model. 
The overall likelihood for the system is then the product of the likelihoods of the individual observed stars:

\begin{equation}
L = \Pi_s (\frac{1}{k}\Sigma{p_{ks}})
\end{equation}

This gives the log likelihoods shown in Table 1. In nearly all cases the extended model has a greater likelihood than the two-burst model, with the 
only exception being Ursa Major I when the stars with questionable membership are excluded.
 
\begin{table}
\centering
\caption{Log likelihoods of the models. The numbers in brackets are those obtained when stars that are now believed not to be members 
of the systems are excluded (J. Simon, private comm.).}
\begin{tabular}{c c c }\\
System&Two-burst&Extended\\
\hline\hline
Bootes I&-12.0&-6.8\\
Canes Venatici II&-20.3 (-14.4)&-4.2 (-3.2)\\
Coma Berenices&-4.9&-3.6\\
Hercules&-14.2&-9.4\\
Leo IV&-3.0&-2.3\\
Ursa Major I&-15.3 (-3.5)&-8.2 (-4.2)\\
\hline
\end{tabular}
\end{table}

\section{Conclusions}

We have modelled the chemical evolution resulting from two possible star formation histories. Our conclusions are as follows:

\begin{enumerate}
\item The two single-age burst model of \citet{brown14} produces too many stars with low [Fe/H], 
with a lag of $\sim0.6~$dex in the model compared to the observations. This could be explained in part by the systems forming at a higher metallicity 
[Fe/H]$\sim-3.5$, rather than $-4$ as assumed in our model. 
 

\item Extended star formation is a better fit to observations of [$\alpha$/Fe] in UFDs 
than the \citet{brown14} two single-age bursts model
for all galaxies except Ursa Major I. The difference is largest for Canes 
Venatici II, in which half the observed stars show suppressed [$\alpha$/Fe], while the \citet{brown14} star formation history has 95\% of stars 
forming in the first burst. 

\item Enrichment within a burst is required to explain the number of low [$\alpha$/Fe] stars in systems where \citet{brown14} determine 
that the vast majority of stars are in the first burst. The success of the Ursa Major I two-burst model suggests that two-burst models can explain at least some 
UFDs, however a more even distribution of stars between the two bursts is required. 
An alternate explanation is extended bursts, such that the first stars in each burst enrich those formed later, either through Type Ia 
supernova enrichment, with a timescale $\gtrsim$100~Myr, or lower mass Type II supernovae ($\gtrsim25$~Myr).

\item Our modelled $M_{\rm{vir}}=10^7$~M$_\odot$ systems with extended star formation can reproduce the observed chemical abundances. This  
provides support to the conclusion of \citet{webster14} that the UFDs can be explained as systems 
with a low formation mass rather than stripping from much larger ($M_{\rm{tot}}\sim10^9$~M$_\odot$) halos.

\end{enumerate}

A combination of isochrone fitting as in \citet{brown14}, 
observations of chemical abundances, and modelling of chemical evolution can reveal the star formation history of UFDs. Knowledge of the star 
formation histories can give insight into star formation processes prior to the epoch of reionization. Deeper spectroscopy with the next 
generation of extremely large telescopes will allow the study of a larger sample of stars, as well as the determination of the 
abundances of more elements. This should allow chemical tagging \citep{bland10,karlsson12}. If stars in dwarf spheroidals and UFDs formed in large clusters, it should be possible to identify stars born within the same cluster, as 
there will be very low scatter 
in chemical abundances between them. \citet{karlsson12} presented tentative evidence for a cluster in 
Sextans at [Fe/H]~$=-2.7$ based on three 
stars with similar [Fe/H] abundances that also had similar Mg, Ti, Cr \& Ba abundances. An equivalent study of UFDs to identify individual clusters 
requires more chemical abundance observations than is currently available. 

\citet{karlsson12} also found that the cumulative metallicity functions of UFDs showed less clustering than for dSphs. They suggest 
that this could result from UFDs forming 
before reionization at masses below the atomic hydrogen cooling limit $M\sim10^8$~M$_\odot$ as in \cite{bovill2009}. Star formation in such 
halos would be affected by inefficient cooling and feedback from Lyman-Werner radiation, which dissociates H$_2$. This could result in a lower 
cluster mass. Lower cluster masses provide support for the result in this paper that UFDs experienced more extended star formation than in the case of large clusters with multiple supernovae.
 
\acknowledgments DW is funded by an Australian Postgraduate Award. JBH is funded by an ARC Laureate Fellowship. We thank the anonymous referee for their useful comments. We thank Josh Simon and David Hogg for discussions relating to this paper.

\end{document}